\documentclass{revtex4-1}
\usepackage{graphicx}
\usepackage{dcolumn}
\usepackage{bm}
\usepackage{tensor}
\usepackage{amssymb}
\usepackage{latexsym}
\usepackage{amsmath}
\newcommand{\G}{\tensor{G}}
\newcommand{\ttau}{\tensor{\tau}}

\newcommand{\T}{\tensor{T}}

\newcommand{\be}{\begin{equation}}
\newcommand{\K}{\tensor{\mathcal{K}}}
\newcommand{\ee}{\end{equation}}
\newcommand{\bea}{\begin{eqnarray}}
\newcommand{\eea}{\end{eqnarray}}

\begin{document}
\title{Perturbations in dense matter relativistic stars induced by
internal sources } 
\author{ Seema Satin \\
Dept. of Physical Sciences, Indian Institute for Science Education and
 Research, Kolkata India }
\email{seemasatin@iiserkol.ac.in}
\begin{abstract}
 Einstein's field equations with a background source term that  
induces perturbations and the applications of this new formalism to a compact 
dense matter relativistic star are presented. We introduce a new  
 response kernel  in the field equations between the metric
 and fluid perturbations. 
A source term which drives or induces the sub-hydro mesoscopic scales
 perturbations in the astrophysical system, is of importance here.
 Deterministic as well as stochastic  perturbations for the radial case are
worked out as solutions of field equations. We also touch upon
polar perturbations  that are deterministic with oscillatory parts.
The stochastic perturbation are of significance in terms of two point or point
 separated correlations which form the building blocks for studying 
equilibrium and 
non-equilibrium statistical  mechanics for the system. Our main aim is to 
build  a theory for intermediate scale physics, for 
 dense exotic matter  
 and investigate structure of the compact astrophysical objects. Specifically,
 turbulence which connects various scales in superfluid matter in the dense
 stars is an area gaining importance. The work presented here  is the
 starting point for new theoretical frame work that  touches upon various
yet unexplored scales where mechanical and dynamical effects interior to
the matter of the star  are of  significance. Thus
there is scope  for extending studies in
asteroseismology to mesoscopic effects at the new intermediate scales
in the cold dense matter fluid . This is expected to enable us to probe 
astrophysical features  at refined scales through theoretical 
formulations as well as for observational consequences.   
We provide a first principles approach with our formulations to study the new
mesoscopic scales which are yet unexplored in asteroseismology and  will form
 a bridge between the macroscopic  scales defined by the hydrodynamics and
microscopic details characterized by nuclear physics and quantum aspects in
the exotic fluid. 
\end{abstract}
\maketitle
\section{Introduction}
Exploring dense matter in compact objects is an active topic of
research  at present \cite{Lattimer, Zdunik,Li, Golam,Nils,Tuhin} and of
 prime importance in asteroseismology . Several
 ongoing efforts to understand the nature
of dense matter focus on microscopic nuclear physics details. There are 
ongoing efforts to  connect the macroscopic picture with the microscopic 
details \cite{Schmitt,Shternin} in order to find out and refine  the equation
 of state which is
still unknown for the interiors and core of the relativistic stars. While
such studies are based on nuclear physics and 
thermodynamics, there are new avenues in the direction of turbulence
and large eddies  in rotating configurations which have come under survey
where magnetohydrodynamics is considered as a rich topic for new investigations 
\cite{Thomas, Thomas2}.
 Mesoscopic studies  with emphasis on relativistic fluids that the 
exotic objects are composed of, have just begun \cite{Seveso,Seveso1}, while 
fluctuations in 
relativistic fluids is another focus area with good scope for new
developments \cite{Mullins}. While the large eddy simulations focus on the
 hydrodynamics and macro scales, an initial attempt for mesoscopic studies
 taking into consideration quantum vortices has just begun \cite{Seveso}.
 We raise an interesting  new query towards sub-hydro mesoscopic scales,
which lie a little below the macroscopic hydro scales but much above the
microscopic and quantum vortex scales. The purpose
of doing so, is to attempt to establish an intermediate scale theory for
 dense compact
matter, where effects of strong gravity can be captured for non-local
properties. For such a study, Einstein's equations in a perturbative approach
 are  important from first principles. With such theoretical
developments,  it will be possible to address long range (or point separated)
  structure inside
 the dense stars and touch upon the scales at which curvature of spacetime 
starts showing effect on the physical variables.   
With such a purpose, recently, a classical Einstein-Langevin  formalism has been
proposed as a new theoretical formulation \cite{Satin1,Satin2}. 
 This has wide scope for theoretical modelling at basic level and 
exploring connections with realistic models of compact stars. Such an effort
 is based on foundational level new constructs in general relativity,
where a stochastic noise term has been added to the linearized Einstein's 
field equations making them inhomogeneous in nature.  
The distinct feature of such a new
construct are the noise term and a reponse kernel, which relates
fluid ( matter) perturbations to the spacetime metric perturbations .   
 While the idea has been borrowed from the semiclassical 
Einstein Langevin equation \cite{Bei}, where noise is due to quantum
 fluctuations, the
classical counterpart has very different construct and applications. A
linear reponse kernel explicity calls for attention towards the new formulation
and needs further elaborate considerations. In the present article we further
modify our earlier attempts,  mathematically, to  better suit  
the astrophysical conditions that we intend to investigate.
In previous work \cite{Satin3} a response function has been taken as a 
bi-scalar, while
here we consider a bi-tensor as a response kernel  that looks more
appropriate, mathematically, as well as for physical considerations in the
 Einstein's equations where component-wise segregation is meaningful
with a metric structure. 
A  consistent noise model that drives the perturbations is also 
important here. Thus we re-build our proposed formalism with a tensor
resonse kernel and a slightly different driving source term. From now on 
we use the term linearized Einstein's equation with a background source instead
of the Einstien-Langevin equation which is restricted to stochastic noise only.
In this article we give a model for deterministic source in the background
spacetime which drives the perturbations, as well as consider a stochastic model
separately on the lines of our previous work. Theoretically these are two
are separate cases,  giving rise to  different
contexts for the perturbations of relativistic stars.   
The simple models of non-rotating configuration  are  
worked out in this article, with analytical results obtained  for radial 
perturbations and the semi-analytical equations for the polar
 perturbations.
We do not however carry out  full solutions for polar perturbations here,
 as that needs seperate rigorous  methods to be 
developed as part of further research with numerical modelling 
for stochasticity  in the Einstein's equations. This will also give scope 
 for new avenues to explore in numerical relativity .  
     
\section{Perturbed Einstein's Equations with a source term}
A linear response relation between metric and fluid perturbations has been
defined in \cite{Satin2}, we modify it slightly in the present article 
with a new form of  response kernel, which is more suitable
for our considerations. The fundamental elements
of perturbation are  $h_{ab} $  for metric, and $\xi^a$ the Lagrangian 
displacement  vector for the matter.  We propose in this article
 a linear response relation between the two  which is of the form 
\be
F^{ab}[h,x) = \int \K{^a^b_c_d}(x-x') M^{cd}[\xi,x')  dx' 
\ee
where $F^{ab}[h]$ represents tensors with metric perturbations 
$h$ and $M^{cd}[\xi]$ denotes tensors with fluid perturbations
$\xi$, while the bitensor $\K{^a^b_c_d}(x-x')$ is the response kernel  
giving a new physically relevant quantity "susceptibility" of spacetime to
 get perturbed.  
We add to this form, an internal  background source which  induces or drives 
 perturbations in the gravitating configuration. The above relation splits
terms of Einstein's equations  into parts with $h_{ab}$ and parts with
$\xi^a$ which can be written phenomenologically, as below. 

 This  form of the linearized Einstein's equations with an added
source as the inhomogeneous part,  reads
\bea \label{eq:els}
\delta \G{^a^b}[h,x) - 8 \pi \delta \T{^a^b}[h,x) - 8 \pi \int 
\K{^a^b_c_d} (x-x') \delta \T{^c^d}[\xi,x') dx' = \ttau{^a^b}[g,x)
\eea
where $\tau^{ab}[g,x) $ is the source of perturbations defined on the
 background $g_{ab}$ and satisfies $\nabla_a \ttau{^a^b}$ w.r.t the background
unperturbed metric . 
 If  the driving source $\tau^{ab}(x) =0 $, we anticipate no induced
 perturbations in the configuration. Hence this formalism is based on the
 inhomogeneous source  term.  

 In this article, we propose two cases for $\tau_{ab}(x)$, 
\begin{itemize}
\item $ \ttau{^a^b} (x) $ defines a deterministic source  in the
background fluid given by $\delta_s \T{^a^b}(\vec{x}) e^{i \omega t}$, 
$\delta_s$
denotes source, as opposed to $\delta $ used in equation (\ref{eq:els})
 for the  perturbed quantities and
\item $ \ttau{^a^b}(x)$ as stochastic or a Langevin source, which is similar to
 the previous formulations in \cite{Satin1,Satin2}, with a slight change. 
We model stochasticity with $\delta_s T^{ab} (\vec{x}, t) e^{i \omega t} $
here $\delta_s$ denotes a stochastic source. Thus the
 amplitude $\delta_s T^{ab} (\vec{x},t) $ in general is
random in space as well as in time, with the oscillatory part given by
 $e^{i \omega t}$ . 
\end{itemize}
In general equation  (\ref{eq:els}) can be seen as a sourced inhomogeneous
 Einstein's perturbed equation rather than the Einstein-Langevin equation as 
proposed in our previous work. 
The response kernel bi-tensor with indices  is the basic
modification to our earlier proposed  linear response kernel for gravitating
systems where it was defined as a bi-scalar. 
 In this article we work out a specific model for such a response kernel which
 looks reasonable for our configuration.
This response kernel gives us a  susceptibility tensor
after  taking a Laplace transform as shown later in the expressions and 
results. 

In the following section, we will give solutions of the equation (\ref{eq:els})
for radial and polar perturbations  of a non-rotating, spherically symmetric
 relativistic star. This is different 
from our earlier work \cite{Satin1,Satin2,Satin3}, in that, the perturbations
 are  
deterministic in  one case, while we also consider  generalized
stochastic nature as the second case. We have   used a more consistent 
source and noise model in this article  compared to previous articles. 

For simplicity of analytical solutions, in the spherically symmetric 
configuration, we assume for the response kernel the form, 
\be \label{eq:rk2}
 \K{^a^b_c_d}(x-x') = \K{^a^b_c_d} (t-t') \delta(r-r')\delta(\theta-\theta')
\delta(\phi - \phi')
\ee
 One can certainly have other models, where the response is felt over
spatial separations intervals also. But that increases the difficulty to get
analytical solutions for the equations and  one will have to resort to
 numerical solutions. 

 With the simple model that we propose here, we intend to present
the basic form of final solutions in closed analytical form which can be
taken as a physically meaningful toy model. 
We  use in the expressions that follow,  the response kernel 
$\K{^a^b_c^d}(x-x')$ , with 3 indices up and one down for convenience such that
the corresponding term in the sourced Einstein's equation, can be evaluated
 using  the form $\int \K{^a^b_c^d}(x-x') \delta \T{^c_d}[\xi,x') dx'$.

In this article, for a perfect fluid matter we  use the  following model
  with non-zero components of the response kernel given by, 
\bea \label{eq:rk1}
& &  \K{^1^1_1^1}(x-x') = \K{^2^2_2^2} (x-x') = \K{^3^3_3^3} (x-x') = 
\bar{\mathcal{K}}_1(x-x') \nonumber \\
& & \K{^0^0_0^0}(x-x') = \bar{\mathcal{K}}_0(x-x'), \K{^0^1_0^1}(x-x') = 
\bar{\mathcal{K}}(x-x')
\eea
Such a model of response kernel is suitable for  perfect fluid perturbations
keeping the response kernel components same for $\delta \T{^1_1},
\delta \T{^2_2} ,\delta \T{^3_3} $ due to pressure perturbations while
$\delta \T{^0_0} $ and $\delta \T{^0_1}$ for energy density and velocity 
variables as different, in general, from that of components for pressure 
variables. 

In the following section, we will solve the sourced Einstein's linearized
equation (\ref{eq:els})  for perturbations (radial and polar) induced in
 the spherically symmetric geometry of a relativistic star, composed of cold
 dense fluid. 
\section{Solutions for radial and non-radial polar perturbations} 
 
A spherically symmetric relativistic star has line element, 
\be
 ds^2 = - e^{2 \nu} dt^2 + e^{2 \lambda} dr^2 + r^2 ( d \theta^2 + 
\sin^2 \theta d \phi^2 )
\ee
 The cold dense matter being modelled as perfect fluid with
\be
 T^{ab}(x) = u^a u^b ( \epsilon + p) + g^{ab} p 
\ee
where
$ u^a u_a =-1 $ and  $ u^a = e^{-\nu}\{ 1,0,0,0 \}$ for the static equilibirum 
case. At a later time a radial velocity $v$ can be introduced in the fluid, 
$v= e^{\lambda - \nu} \dot{r}$ such that four-velocity has components,
$u^a \equiv \frac{1}{\sqrt{1-v^2}} (e^{- \nu}, e^{-\lambda} v, 0,0) $.

Accordingly, the components of field equations are given as
\bea
G^t_t = 8 \pi T^t_t & : & \nonumber \\
  e^{-2 \lambda} (\frac{1}{r^2} - \frac{2}{r}
\lambda ' )& - & \frac{1}{r^2} = - 8 \pi \epsilon \frac{1}{1-v^2} \label{eq:1}\\
G^r_r = 8 \pi T^r_r & : & \nonumber \\
e^{-2 \lambda}(\frac{1}{r^2} + \frac{2}{r} \nu' ) & - & \frac{1}{r^2} = 8
\pi ( \epsilon \frac{v^2}{1-v^2} + p ) \label{eq:2} \\
G^t_r = 8 \pi T^t_r & : & \nonumber \\
- \frac{2}{r} e^{-2 \nu} \dot{\lambda}  & = & 8 \pi e^{\lambda - \nu}
(\epsilon + p) \frac{v}{1- v^2} \label{eq:3}\\
e^{ 2 \lambda} G^\theta_\theta = 8 \pi e^{2 \lambda}  T^\theta_\theta
& : & \nonumber \\
\nu '' + \nu'^2 & - & \nu ' \lambda ' +  \frac{1}{r} (\nu ' - \lambda ' )
= 8 \pi e^{2 \lambda} p
\eea
Also we can easily obtain the relation,
\be
\nu' + \lambda' = 4 \pi (\epsilon + p ) e^{2 \lambda} p
\ee
from the above Einstein's equations. Its easy to see , if we put $v=0$, in the
above, we get equations for the static configuration. The radial
velocity has been introduced
to get the correct form of perturbed field equations in a heuristic way
, such that after
perturbing the equations correctly, one can put $v=0$ back in the
unperturbed equations. 
 The perturbations in the fluid can be
introduced using the radial velocity such that,
\be
\delta v = e^{\lambda - \nu} \dot{\xi}
\ee
where $\xi \equiv \xi_r$ is the only non-zero fluid displacement vector.

\subsection{Radial perturbations induced by the oscillating sources }
In this section we work out solutions for radial perturbations near the
end stages
 of collapse of a relativistic star. Ideally during a core collapse, one gets
a rotating configuration, which will need numerical solutions for perturbed
field equations. Our
purpose is to lay foundations for the study of perturbations which are
 induced  due to internal sources present in the fluid at mesoscopic
scales. The collapse mechanism itself can give rise to mesoscopic oscillations
inside the core of dense matter, where the degeneracy pressure and energy 
density fluctuations act as seeds that can grow for different modes of 
oscillations. One such internal effect is the turbulence in the interior of the 
cold dense matter fluid, which has not yet been modelled so far from the
first principles. For  perturbations induced by such effects
we restrict the present model in order to account for analytical expressions
 and hence consider a non-rotating model. The   analytical
form give us better insight to further implement 
numerical solutions for more involved and realistic cases . We
expect to carry out this in future work.  It is  important to formulate
the new theoretical base with simpler  models first.   

\subsubsection{Models of source term in cold dense matter} 
 The cold dense matter is modelled by a perfect fluid. In this article we
consider two models, $(i)$  deterministic  oscillations 
$\delta_s T^{ab} (\vec{x}) e^{i \omega t} $ of the stress tensor, and $(ii)$
 stochastic fluctuations of the stress tensor $\delta T^{ab}(\vec{x},t) e^{i 
\omega t} $. 

$(i)$ Thus $\ttau{^a^b}(x) $ has the following non-zero components for the
deterministic case.
\bea \ttau{^0_0}(r,t) &=& \delta_s \epsilon (r) e^{i \omega t} 
\nonumber \\
\ttau{^1_0}(r,t) &=&  (\epsilon(r)+p(r)) \delta_s v(r) e^{ i \omega t} \nonumber \\
\ttau{^1_1}(r,t) & = & \delta_s p(r) e^{i \omega t} \nonumber \\
\ttau{^2_2} (r,t) & = & \delta_s p(r) e^{i \omega t} \nonumber \\
\ttau{^3_3} (r,t) & = & \delta_s p(r) e^{i \omega t} \nonumber \\
\eea   
Here 's' in $\delta_s $ denotes source, and the radial amplitudes  oscillate
with frequency $\omega$. Such oscillations can be taken due to the collapse 
mechanism itself which leaves  the fluid oscillating in the interiors. Due
to internal dynamics of the configuration then, these can grow and affect 
 the linear perturbations in many ways. As we have mentioned earlier such
end of  collapse states are dynamical and rotating, it leaves  substantial
scope for dynamical effects to take over at mesoscopic scales and affect the 
perturbations of the  gravitating system. We touch upon with our formulations
the coarse grained effects.

(ii) The stochastic fluctuations of $\tau^{ab}(x) $ have the following
 non-zero components,
\bea \label{eq:sto1}
\ttau{^0_0}(r,t) &=& \delta_s \epsilon (r,t) e^{i \omega t} \nonumber \\
\ttau{^1_0}(r,t) &=&  (\epsilon(r)+p(r)) \delta_s v(r,t) e^{ i \omega t} 
\nonumber \\
\ttau{^1_1}(r,t) & = & \delta_s p(r,t) e^{i \omega t} \nonumber \\
\ttau{^2_2} (r,t) & = & \delta_s p(r,t) e^{i \omega t} \nonumber \\
\ttau{^3_3 }(r,t) & = & \delta_s p(r,t) e^{i \omega t} \nonumber \\
\eea   
 Here $\delta_s v, \delta_s p,\delta_s \epsilon$ are defined stochastically
and hence have meaning only as statistical averages given by
 $ \langle \delta_s v \rangle, \langle \delta_s \epsilon \rangle , \langle
\delta_s p \rangle $.  The expectation is taken with respect to
both space  and time variables.  However we see that the amplitudes
 of oscillations in the stochastic model are functions of the temporal
variable also, though we have emphasized more on the generalized stochastic
models in our earlier work \cite{Satin2}.

In  general, the  sources are defined such that 
  , for example, $ \delta_s p(r,t) = p(r,t) - \langle p(r,t) \rangle $ , where
$\langle p(r,t) \rangle $ is the expectation or average  of the pressure in the
background unperturbed configuration. For a Langevin noise $\langle 
\ttau{^a^b} \rangle = 0 $ hence one needs to obtain the rms value or two
point correlations for a meaningful analysis. 
\subsubsection{The radial perturbations }
There should be no confusion between the  sources and the perturbations in the
 system. While sources are the fluctuations in the background and denoted 
by $\delta_s$, perturbations are defined as shift in the trajectories and
the metric given by $h_{ab} $ and $\xi^a$. 

For near equilibrium configuration, consider $\delta v_r(r,t), \delta p(r,t),
\delta \epsilon(r,t) $ as radial perturbations in the fluid   where $\xi \equiv
\xi_r $   is the only non-zero component of the Lagrangian displacement
vector, then, 
\bea
\delta p[\xi] & = & - \Gamma_1 p\frac{e^{-\lambda}}{r^2} [ e^{\lambda} r^2
\xi ]' - \xi p' \\
\delta \epsilon [\xi] & = & - (p + \epsilon)
\frac{ e^{-\lambda}}{r^2} [ e^{\lambda} r^2 \xi]' - \xi \epsilon' \\
\delta u^r[\xi] & = &  \mathcal{L}_u \xi^r = e^{-\nu} \dot{\xi}
\eea
 The remaining part $\delta p[h] = - \Gamma_1 p \delta \lambda$ and 
$\delta \epsilon[h]= -(p + \epsilon)\delta \lambda $, while $\delta u^r[h] =0$. 
We assume for a near-equilibrium spherically symmetric configuration, the 
 non-zero components (\ref{eq:els}) to read,
\bea
& & \delta \G{^0_0}[h;x) - 8 \pi \delta \T{^0_0}[h;x) - 8 \pi \int 
\bar{\mathcal{K}}_0 (x-x')
\delta \T{^0_0}[\xi,x') dx' = \ttau{^0_0}[g;x) \label{eq:tt} \\
& & \delta \G{^0_1}[h;x) - 8 \pi \delta T^0_1[h;x) - 8 \pi \int 
\bar{\mathcal{K}}(x-x')
 \delta \T{^0_1}[\xi,x') dx' = \ttau{^0_1}[g;x) \label{eq:tr} \\
 & & \delta \G{^1_1}[h;x) - 8 \pi \delta \T{^1_1}[h;x) - 8 \pi \int 
\bar{\mathcal{K}}_1(x-x') \delta \T{^1_1}[\xi,x')  dx' = \ttau{^1_1}[g;x)
 \label{eq:rr} \\
& & \delta \G{^2_2}[h;x) - 8 \pi \delta \T{^2_2}[h;x) - 8 \pi \int 
\bar{\mathcal{K}}_1 (x-x') \delta \T{^2_2}[\xi,x') dx' = \ttau{^2_2}[g;x)
 \label{eq:thth} \\
& & \delta \G{^3_3}[h;x) - 8 \pi \delta \T{^3_3}[h;x) - 8 \pi \int 
\bar{\mathcal{K}}_1(x-x') \delta \T{^3_3}[\xi,x') dx' = \ttau{^3_3}[g;x)
 \label{eq:phph}
\eea
where we have used the forms of response kernel tensor given in equation
(\ref{eq:rk1}).
\subsubsection{Solutions for deterministic induced perturbations in cold dense
 matter}  \label{sec:sr}
To obtain solutions from equations (\ref{eq:tt}), (\ref{eq:tr}),
(\ref{eq:rr}) we assume the form of $\xi(r,t) \equiv \xi(r) e^{ \gamma_r t}$.
Then, from equation (\ref{eq:tr}) and the source term with $ \delta_s v(r,t)
 = \delta_s v(r) e^{i \omega t}$ along with the response kernel from
 (\ref{eq:rk2}) and (\ref{eq:rk1}), we have, 
\be \label{eq:lam1}
\delta \lambda(r,t) = (\nu' + \lambda')  \bar{\mathcal{K}}(\gamma_r) 
\xi(r,t) + i e^{(\nu - \lambda)} (\nu' + \lambda')  \frac{\delta_sv(r)
 e^{i \omega t}}{ \omega} 
\ee
where Laplace transform for the response kernel is given by
 $\bar{\mathcal{K}}(\gamma_r)  = \int \mathcal{K}_0 (\tau) 
e^{-\gamma_r \tau} d\tau $, $\tau = t-t'$ has been used. 

Substituting this in equation (\ref{eq:tt}),
\be \label{eq:xi}
g(r) \xi'(r,t) + f(r) \xi(r,t) =  -  j(r)  i \delta_s v(r) \frac{e^{i 
\omega t}}{\omega} - 8 \pi \delta_s \epsilon(r,t)
\ee
where
\bea
 g(r) & = & - \frac{(\nu'+\lambda')}{r} [2 \bar{\mathcal{K}}(\gamma_r) -
 \bar{\mathcal{K}}_0 (\gamma_r)]  \\
f(r) &= & -[\frac{2}{r} \{(\nu' + \lambda') \bar{\mathcal{K}}(\gamma_r) \}'
- (\frac{3 \lambda'}{r} - \frac{2}{r^2} - \frac{\nu'}{r} ) (\nu' + \lambda')
\bar{\mathcal{K}}(\gamma_r) - \frac{(\nu' + \lambda')}{2} (\lambda' +
 \frac{2}{r} ) + 8 \pi \epsilon' e^{2 \lambda} ] \\
 j(r) & = & ( \frac{3 \lambda'}{r} - \frac{2}{r^2} - \frac{\nu'}{r}) e^{\nu - 
\lambda} ( \nu' + \lambda') 
\eea
 Solution of equation (\ref{eq:xi}) gives,
\bea \label{eq:solxi}
\xi(r,t) & = & e^{\int \frac{f(r')}{g(r')}dr'} \int e^{- \int 
\frac{f(r'')}{g(r'')} dr'' } ( j(r') i \frac{\delta_s v(r') e^{i \omega t}}{
\omega} - 8 \pi \delta_s \epsilon(r',t) ) dr' 
\eea
thus from (\ref{eq:lam1}) we get 
\bea \label{eq:solam}
\delta \lambda (r,t) & = & (\nu\ + \lambda') \bar{\mathcal{K}}(\gamma_r) 
e^{\int \frac{f(r')}{g(r')} dr' } \int e^{- \int \frac{f(r'')}{g(r'')} dr''}
[\{ j(r') - e^{\nu(r') - \lambda(r')} ( \nu' (r)+ \lambda'(r) ) \delta (r-r')\}
 \nonumber \\
& &  ( i \frac{\delta_s v(r') e^{i \omega t}}{\omega})  - 8 \pi 
  \delta \epsilon_s(r',t)] dr' 
\eea

From equation (\ref{eq:rr}),
\bea \label{eq:solnu}
\delta \nu(r,t) & =  & \int \int \{ [\mathcal{L}_1(r'') g_1(r') +
 \mathcal{L}_2(r'') g_2(r')
+ \mathcal{L}_3(r') g_3(r') ] (i \frac{\delta_s v(r')e^{i \omega t}}{\omega})
  \nonumber\\
& & 
- 8 \pi  \bar{\mathcal{K}}(\gamma_r) e^{m_1(r')} [ (\mathcal{L}_4 (r'') + 
\mathcal{L}_6 (r'') + \mathcal{L}_5 (r'') ( 1+ \delta(r-r') ] \delta_s
\epsilon(r',t)  \} dr' dr''  \nonumber \\
& & + 8 \pi \int \delta_s p (r',t) dr' 
\eea
where
\bea
m_1(r) & = & \int f(r')/g(r') dr' \nonumber \\
 \mathcal{L}_1(r) & = & \{2 e^{-2 \lambda} ( \frac{1}{r^2} + \frac{2}{r} \nu' )
 - 8 \pi \Gamma_1 p \} \{ (\nu'+ \lambda') \bar{K}(\gamma_r) e^{- m_1(r)} \}
\nonumber \\
g_1(r) & = & ( j(r) - e^{\nu(r) - \lambda(r)} (\nu'(r') - \lambda'(r')) 
\delta(r-r') ) e^{m_1(r)} \nonumber \\
\mathcal{L}_2 (r) & = & 8 \pi \bar{\mathcal{K}}_1(\gamma_r) \Gamma_1 p m_1(r) 
e^{-m_1(r)} \nonumber \\
g_2(r) & = & e^{m_1(r)} ( 1+ \delta(r-r')) j(r') \nonumber \\
\mathcal{L}_3(r) & = &  \bar{\mathcal{K}} (\gamma_r) ( \frac{ \gamma_1 p
 e^{- \lambda}}{r^2} (e^\lambda r^2)' + p' ) e^{- m_1(r)} \nonumber \\
g_3(r) & = & e^{m_1(r)} j(r) \nonumber \\
\mathcal{L}_4(r) & = & ( \nu'+ \lambda') \{ 2 e^{-2 \lambda} ( \frac{1}{r^2} 
+ \frac{2}{r} \nu' ) - 8 \pi \Gamma_1 p \} e^{- m_1(r)} \nonumber \\
\mathcal{L}_5(r) & = & 8 \pi \bar{\mathcal{K}}(\gamma_r) m_1'(r) e^{-m_1(r)}
\nonumber \\
\mathcal{L}_6 (r) & = &  (\Gamma_1 p e^{- \lambda} ( e^\lambda r^2)' + p')
e^{- m_1(r)}  
\eea
 
 Equations (\ref{eq:solxi}), (\ref{eq:solam}) , (\ref{eq:solnu}) are
the main results for the deterministic perturbations. One can see
 that the rhs of these equations consists of the integrated effect of the 
sources $\delta_s v(r,t) ,\delta_s \epsilon (r,t), \delta_s p(r,t) $ over some
radial depth.
It is hence important to take care of the the causal effects, for the 
limits of integration in the expressions, while doing numerics and evaluating 
the terms explicitly. For example a relevant condition can be put as
the radial depth  $r-r' \leq c_s(t-t')$ where $c_s$ is the velocity of 
sound in the interior of the fluid. Interestingly we see $\delta \lambda,
\delta \nu, \xi_r$ have explicit imaginary parts and are complex in nature also
for the amplitudes of the oscillations. we will do further analysis of these
perturbations in the work to follow in future article where we will build up
more on mesoscopic theory for the exotic matter. The expressions presented here
will be used as building blocks. 
\subsection{Stochastic perturbations in cold dense matter }
For  modeling stochastic effects, we consider the noise model given by
equation (\ref{eq:sto1}) in the earlier subsection. The stochastic
perturbations are of the form $\xi(r,t) \equiv \tilde{\xi}(r,t) e^{\gamma_r t}$
where $\gamma_r$ is complex in general and the amplitude $\tilde{\xi}(r,t)$
is a random function of $t$. For a generalized  stochastic nature as defined in
\cite{Satin2}( randomness w.r.t spatial as well as temporal coordinates 
for a spacetime structure), we also assume the randomness w.r.t '$r$' . Here we
 consider the stochastic noise model $(ii)$ for $\ttau{^a^b}$.   

Then  from (\ref{eq:tr}) we obtain,
\bea
& & \dot{\delta \lambda} + (\nu'+\lambda') \int \bar{\mathcal{K}}(t-t')
 \dot{\xi}(r,t') dt' = - e^{\lambda - \nu}(\lambda' + \nu') \delta_s v(r,t)
\eea
 The term with integral over the response kernel can be written as,
\[
 \int \bar{\mathcal{K}} (\tau) \dot{\xi}(r,t-\tau) d \tau
\]
using Taylor expansion for $\dot{\xi}(r,t-\tau)$ we obtain,
\bea \label{eq:tr2}
\delta \lambda = (\nu'+ \lambda') ( \tilde{\xi}(r,t) \bar{\mathcal{K}} 
(\gamma)- \dot{\tilde{\xi}}(r,t)
\tilde{\mathcal{K}} (\gamma) ) e^{\gamma_r t} - e^{\lambda - \nu} (\lambda' +
 \nu') \int \delta_s v(r,t') dt'
\eea
where $\bar{\mathcal{K}}(\gamma_r) = \int \bar{\mathcal{K}}(\tau) e^{-\gamma_r 
\tau} d\tau$ with $\tau = t-t'$, and $\tilde{\mathcal{K}}(\tau) = \tau 
\bar{\mathcal{K}}(\tau)$ such that $\tilde{\mathcal{K}} (\gamma_r)  = \int
\tilde{\mathcal{K}}(\tau) d \tau $. Similar relations hold for 
$\bar{\mathcal{K}}_0(\gamma_r), \bar{\mathcal{K}}_1(\gamma_r), 
\tilde{\mathcal{K}}_0(\gamma_r) $ and $\tilde{\mathcal{K}}_1(\gamma_r)$.
 From equation (\ref{eq:tt}),one gets
\bea \label{eq:tt2}
& & - 2 e^{- 2 \lambda} \frac{\delta \lambda}{r} ( \frac{1}{r} \nu' - \lambda')
- \frac{2}{r} \delta \lambda' e^{-2 \lambda}  8 \pi 
\int \{\bar{\mathcal{K}}_0 (t-t') \{(\epsilon + p ) \xi'(r,t') \nonumber\\
& &  + (\lambda' + \frac{2}{r}+\epsilon'(r)) \xi(r,t') \} dt' = 
8 \pi \delta_s \epsilon(r,t)
\eea
 From equation (\ref{eq:rr}),  the following can be obtained,
\bea
& & 2 e^{-2 \lambda}[ \frac{\delta \nu'}{r} + ( \frac{1}{r^2} + 
\frac{2 \nu'}{r}) \delta \lambda ] + 8 \pi \Gamma_1 p \delta \lambda 
 - 8 \pi \int \bar{\mathcal{K}}_1(t-t') \Gamma_1 p [ \xi'(r,t') + 
(\frac{2}{r} + \nonumber \\
& &  \lambda' + \frac{\epsilon'}{\epsilon + p} ) \xi(r,t') ]= \delta_s p(r,t)
\eea
We also use the covariant conservations of the stress tensor and perturb it
\be
 \delta \nabla_a \T{^a_b}(x) = 0 
\ee
The component $\delta \nabla_a \T{^a_0}(x) =0 $ gives,
\bea \label{eq:cov0}
 & & - 2 (\epsilon+p) \delta \lambda + \xi [ - (\epsilon+p)(\lambda' +
 \frac{2}{r}) - (p' + \lambda' + \nu' + \frac{1}{2 r} ) ] = 0 
\eea
 From (\ref{eq:tr2}) and (\ref{eq:cov0}),
\bea
\tilde{\xi}(r,t) e^{\gamma_r t} & = & e^{(\gamma_r- X_1(r))t} \int e^{(X_1(r) -
 \gamma_r) t''} x_1(r) (\int \delta_s v(r,t') dt') dt''
\eea
where 
\be
X_1(r) = \frac{1}{ \tilde{\mathcal{K}}(\gamma_r)} [ (\nu' + \lambda') 
\bar{\mathcal{K}}(\gamma_r) + \frac{1}{2} (\lambda' + \frac{2}{r}) +
 \frac{1}{2} \frac{( p' + \lambda' + \nu' +1/(2 r) )}{(\epsilon+p)}] 
\ee 
and 
\be
x_1(r) = - \frac{e^{\lambda - \nu}}{\tilde{\mathcal{K}}(\gamma_r)}
 (\lambda' + \nu')
\ee

\bea
\delta \lambda & = & (\nu'+ \lambda') [ (\bar{\mathcal{K}}(\gamma_r) - X_1(r))
 e^{(\gamma_r - X_1(r)) t } \int e^{(X_1(r)- \gamma_r) t'' } x_1(r) \int 
\delta_s v(r,t') dt' dt''] \nonumber \\
& & - (\nu' + \lambda') (\tilde{\mathcal{K}}(\gamma_r) x_1 (r) +
 e^{\lambda - \nu} ) (\int \delta_s v(r,t')dt') 
\eea
 Substituting  
\bea
\delta \nu(r,t) & = & \int \mathcal{W}(r',t,t'') \delta_s v(r',t') dt'
dt'' dr' - 8 \pi \int \delta_s p(r',t)dr'
\eea
where 
\bea
\mathcal{W}(r',t,t'') & = & S_1(r') [ x_1(r')(1- X_1(r'))
 e^{(\gamma_r - X-1(r'))t + (X_1(r') - \gamma_r)t''}] + S_2(r')[ \{x_1'(r')
(1+ X_1(r')) + \nonumber \\
& & x_1(r')(X_1(r')X_1'(r')(t-t'')- X_1'(r') ) \}
e^{(\gamma_r - X_1(r)) + (X_1(r')- \gamma_r)t''}]  \nonumber \\
 & &- S_3(r') [ (X_1'(r') t + x_1'(r') - x_1(r')(1+X_1'(r') t'') )
 e^{- X_1(r') t + (X_1(r')- \gamma_r)t'' }] 
+  \nonumber \\
& & S_4 (r') [ e^{- X_1(r') t + (X_1(r') 
  - \gamma_r ) t'' } x_1(r') ] + S_5(r') 
\eea
where
\bea
S_1(r) & = & \frac{e^{2 \lambda}}{2} r [- \tilde{\mathcal{K}}(\gamma_r)
(\nu' + \lambda') \{ 8 \pi \Gamma_1 p - 2 e^{-2 \lambda} (\frac{1}{r^2} + 
\frac{2}{r}\nu' ) \} - 8 \pi ( \Gamma_1 p (\lambda' + \frac{2}{r}) + p')
\tilde{\mathcal{K}}_1(\gamma_r) ] \nonumber \\
S_2(r) & = & e^{2 \lambda} 4 \pi r p\tilde{\mathcal{K}}_0 ( \gamma_r) 
\nonumber \\
S_3 (r) & = & 4 \pi\Gamma_1 p r \tilde{\mathcal{K}}_1 e^{2 \lambda}
 \nonumber\\ 
S_4 (r) & = & ( \nu'' + \lambda'') \{ 8 \pi  \Gamma_1 p - 2 e^{-2 \lambda}(
\frac{1}{r^2} + \frac{2}{r} \nu' ) \} \tilde{\mathcal{K}}(r) \frac{r}{2}
e^{2 \lambda} + 4 \pi r e^{-2 \lambda} (\Gamma_1 p( \lambda'+ \frac{2}{r}) + 
p' ) \tilde{\mathcal{K}}_1(r) \nonumber \\
S_5 (r) & = & e^{3 \lambda -\nu} r ( \nu' + \lambda') \{ 4 \pi \Gamma_1 p- 
e^{- 2 \lambda} (\frac{1}{r^2} + \frac{2}{r} \nu' ) \} 
\eea
 
In the expressions  obtained above $\delta_s v(r,t), \delta_s \epsilon(r,t)
$ and $\delta_s p(r,t)$ are of the form $\delta_s s(r,t) \equiv 
\delta_s \tilde{s}(r,t) e^{i \omega t}$ where the amplitude
$\tilde{s}(r,t)$ is stochastic in nature and with oscillatory part
 $e^{i \omega t}$. Thus the above equations are stochastic in
nature, this only the expectation value is meaningful. We assume Langevin 
type noise, hence $\langle s(r,t) \rangle =0 $ while the two point 
correlations are given by $\langle s(r_1,t_1)  s(r_2, t_2) \rangle $.  
Thus,  it can be easily seen that $\langle \xi(r,t) \rangle = 0, 
\langle \delta \lambda(r,t) \rangle =0, \langle \delta \nu(r,t)  \rangle 
=0 $, while the two point correlations take  the form 
\bea
\langle \xi^*(r_1,t_1) \xi(r_2,t_2)\rangle & = & 
 e^{(\gamma_{r_1}^*- X_1^*(r_1))t_1  + (\gamma_{r_2} - X_1(r_2))t_2} t_1 t_2
\int  e^{(X_1^*(r_1) - \gamma_{r_1}^*)t_1'' + (X_1(r_2) - \gamma_{r_2})t_2''}
x_1^*(r_1) x_1(r_2) \nonumber \\
& &  \langle \delta_s v^*(r_1,t_1') \delta_s v(r_2, t_2') \rangle
 dt_1' dt_2' dt_1'' dt_2''
\eea

\bea
\langle \delta \lambda^*(r_1,t_1) \delta \lambda(r_2, t_2) \rangle & = & 
(\nu'(r_1) + \lambda'(r_1)) ( \nu'(r_2) + \lambda'(r_2)) 
[ \bar{\mathcal{K}}^*(\gamma_{r_1}) - X_1^*(r_1))(\bar{\mathcal{K}}
(\gamma_{r_2}) - X_1(r_2)) \nonumber \\
& &  e^{(\gamma_{r_1}^* - X^*(r_1) ) t_1 + (\gamma_{r_2}  - X_1(r_2))t_2}
\int  e^{X_1^*(r_1) - \gamma_{r_1}^*) t_1'' + ( X_1(r_2) - \gamma_{r_2})
t_2''} x_1^*(r_1) x_1(r_2) \nonumber\\
& &  \langle \delta_s v^*(r_1,t_1') \delta_s v(r_2, t_2') 
\rangle dt_1' dt_2' dt_1'' dt_2'' 
  - (\nu'(r_1) + \lambda'(r_1))(\nu'(r_2) + \lambda'(r_2)) 
 ( \tilde{\mathcal{K}}^*(\gamma_{r_1}) x_1^*(r_1)  \nonumber \\
& & + e^{\lambda(r_1)
-\nu(r_1)} ) ( \tilde{\mathcal{K}}(\gamma_{r_2}) x_1(r_2) + e^{\lambda(r_2) -
 \nu(r_2)})
\int \langle \delta_s v^*(r_1,t_1') \delta_s v(r_2,t_2') \rangle dt_1' dt_2'
\eea
\bea
\langle \delta \nu^*(r_1,t_1) \delta \nu(r_2,t_2) \rangle & = & \int 
\mathcal{W}^*(r_1',t_1,t_1'') \mathcal{W}(r_2',t_2',t_2'') 
\langle \delta_s v^*(r_1',t_1')  \delta_s v(r_2',t_2') \rangle dt_1' dt_2'' 
dr_1' + \nonumber \\
& & 16 \pi^2 \int \langle \delta_s p^*(r_1',t_1) \delta_s p(r_2',t_2) \rangle
dr_1' dr_2'
\eea
Similar relation for $\langle \delta \nu^*(r_1,t_1) \delta \lambda(r_1,t_1) 
\rangle $ etc can also be written.
 The point separated correlations in the coincident limit
can be used to find out root mean square
of the perturbations that give an estimate of their magnitudes. 
However our main interest here lies in the the two point or point separated form
 of the expressions which will act as the building block to study extended 
properties of the  compact configuration. The two points can be separated for 
a large distance (keeping causality  condition in mind ) such that correlations
probe extended structures inside the dense fluid matter giving access to non-
local properties and phenomena. Further discussion of results is given in the
concluding section. 

In the next section we give results for polar perturbations in the non-rotating
configuration, 
\subsection{Polar Perturbations}
In this section we solve the sourced Einstein's equations for 
polar perturbations. The induced 
perturbations may be deterministic or stochastic depending on the 
   source . Similar to  the case of
 radial perturbations we can have these two possibilities here. We work out the
deterministic induced perturbations in this article for the polar perturbations,
stochastic polar perturbations can be worked out in an analogous way. For a 
non-rotating geometry we  proceed as follows. 

\subsubsection{ Modelling source term for induced polar perturbations}
We will present the polar perturbatons in Regge-Wheeler gauge.
The  source term $\ttau{^a^b}$  can
be shown to have polar decomposition of  $\ttau{_a_b} $ as a pull back on the
  2-sphere which takes the form
\be
\ttau{_i_j} = e_{ij} \tau_{lm}^{scalar} Y_{lm}+  \tau_{lm}^{polar}
 \nabla_i \nabla_j Y_{lm}
\ee
where $e_{ij}$ is the metric restricted to the 2-sphere.
The indices $i,j $ are reserved for $\theta, \phi $, while $a, b $ run
 for $\{t,r,\theta, \phi \} $.
In the Regge-Wheeler gauge $\tau_{lm}^{polar} =0 $ corresponding to the
decomposition of the metric (see appendix).
The projection for $t-t$ and $r-r$ and $t-r$ components on the 2-sphere can be
 given by,
\bea
& & \ttau{^t_t}_{lm} Y_{lm} =  \delta_s \epsilon_{lm}(r,t) Y_{lm}
(\theta,\phi)  ; \ttau{^r_r}_{lm} Y_{lm} = 
 \delta_s p_{lm} (r,t)  Y_{lm}(\theta,\phi) ;\nonumber\\
& & \ttau{^t_r}_{lm} = e^{\nu - \lambda}( \epsilon+ p) \delta_s v_{lm}(r,t) 
Y_{lm}
\eea
The $\theta,\phi$ components have the form,
\bea
& &  \ttau{^\theta_\theta}_{lm}(r,t) Y_{lm}(\theta,\phi) =
\delta_s p_{lm}(r,t) Y_{lm}(\theta,\phi)  \nonumber \\ 
& & \ttau{^\phi_\phi}_{lm} (r,t) Y_{lm}(\theta,\phi)  =  
 \delta_s p_{lm}(r,t) Y_{lm}(\theta,\phi)
\eea
As we see further, these form the sources for induced perturbations .
Evaluating in this article for $m=0$ case, the non-zero components of the 
source are given as,
\bea
& & \ttau{^t_t} =  \delta_s \epsilon_l(r,t) P_l(\cos \theta)  ;
\ttau{^r_r} =  \delta_s p_l(r,t) P_l(\cos \theta); 
\ttau{^t_r} = e^{\nu - \lambda}( \epsilon+ p) \delta_s v_l(r,t)
 P_l(\cos \theta) \label{eq:tt1}\\
& & \ttau{^\theta_\theta }=  \delta_s p_l(r,t)  P_l(\cos \theta)  
 \nonumber \\
& & \ttau{^\phi_\phi} = \delta_s p_l(r,t) P_l (\cos \theta) 
\eea
 We consider in this subsection $\delta_s p, \delta_s \epsilon $ and $
\delta_s v$ of the form $\delta_s s(r) e^{i \omega t} $, which are
deterministic sources. Similar results can be obtained for 
stochastic noise. 
 We have worked out similar results for the
Einstein-Langevin equation in \cite{Satin3} as a stochastic case, and with a
 different response kernel. Our results in the article
 are based on the Einstein's sourced equation with the specific and 
more resonable response kernel, and are presented as deterministic
 oscillations. These results will have significance in exploring mesoscopic
scale effects in the interiors of dense matter for various new aspects in
rheological modelling of the exotic fluids. 
\subsubsection{The sourced Einstein's equations for polar perturbations}
The sourced Einstein's equation for the polar perturbations, borrows the 
standard notation for the  perturbed quantities in the field equation, with
the new response  kernel and source  term incorporated according to
the proposed equation (\ref{eq:els}).The response kernel is  
of the same form as in subsection (\ref{sec:sr}), and we also
assume the perturbations below,  $H_0, H_1,K, W$ and $V $ are of the form 
$\mathcal{S}(r) e^{\gamma_r t}$.   
 
We have used the standard form ( as given in appendix to denote the
perturbations) of metric and fluid perturbations.  Writing,   
the non-zero components of the  sourced field equations as follows,  

The $t-t$ component is given by,
\be
\delta \G{^t_t} [h,x) -8 \pi  \delta \T{^t_t}[h,x) - 8 \pi \int 
\mathcal{K}_0(x-x') \delta \T{^t_t} [\xi,x') d^4x' = \ttau{^t_t}[g,x)
\ee
which with the polar decomposition reads,
\bea \label{eq:tt1}
 & & [ \{- e^{-2 \lambda} \partial_r^2 - e^{-2 \lambda} ( \frac{3}{r} -
 \lambda') \partial_r + [ \frac{1}{2 r^2} (l-1)(l+2) + 8 \pi (\epsilon+p)]
 \} K +\{ \frac{e^{-2 \lambda}}{r} \partial_r - e^{-2 \lambda} \nonumber \\
& & ( \frac{1}{r^2} - \frac{2}{r} \lambda ' + \frac{e^{2 \lambda}}{ 2 r^2}
 l(l+1)) + 4 \pi (\epsilon + p) \} H_0 +  \tilde{\mathcal{K}}_0(\gamma) 8 \pi
(\frac{e^{-\lambda}}{r^2} ( (\epsilon+p)\partial_r + \epsilon') W
+  (\epsilon+p) \frac{l(l+1)}{r^2} V) ] P_l (\cos \theta)= \nonumber  \\
& & -\delta_s
\epsilon_l(r) e^{i \omega t} P_{\tilde{l}}(\cos \theta)
\eea
The $t-r$ component reads
\be
\delta \G{^t_r} [h,x) -8 \pi  \delta \T{^t_r}[h;\xi, x)   =
\ttau{^t_r}(x) 
\ee
given by
\be \label{eq:tr1}
[e^{- \lambda + \nu} (\partial_r + 2 \nu') H_0 - e^{-\lambda + \nu}
 \partial_r K + \mathcal{K}(\gamma) (\epsilon+ p) \frac{e^{-2 \nu + 
\lambda}}{r^2} \partial_t W] P_l(\cos \theta) =
 8 \pi e^{(\nu-\lambda)} (\epsilon+ p) \delta_s v_l(r,t)
P_{\bar{l}}(\cos \theta) 
\ee
The $r-r$ component is given by,
\be
\delta \G{^r_r} [h,x) -8 \pi  \delta \T{^r_r}[h,x) - 8 \pi \int \mathcal{K}_1 
(x-x') \delta \T{^r_r} [\xi,x') d^4x' = \ttau{^r_r}[g,x)
\ee
with its projection on the two sphere reads,
\bea \label{eq:rr1}
& & [ \{e^{-2 \nu} \partial_t^2 + e^{-2 \lambda} (\frac{1}{r} - \nu')
\partial_r - 8 \pi \Gamma_1 p + \frac{1}{r} (l-1)(l-2) \} K  -
\{ \frac{e^{-2 \lambda}}{r} + \frac{1}{r^2} ( e^{-2 \lambda} -1)
 + \frac{1}{r} (l-1)(l+2) +  \nonumber\\
& & 4 \pi \Gamma_1 p \} H_0 - 8 \pi \tilde{\mathcal{K}}_1 (\gamma) [ \Gamma_1 p
 \frac{e^{-\lambda}}{r^2} \partial_r + \frac{p'}{r^2} e^{-\lambda}] W -
 8 \pi \tilde{\mathcal{K}}(\gamma) \Gamma_1 p \frac{l(l+1)}{r^2} V ]
P_l (\cos \theta) = \delta_s p_{\tilde{l}}(r) e^{i \omega t} P_{\tilde{l}}
 (\cos \theta).
\eea

The $\theta-\theta $ component is given by,
\be \label{eq:the11}
\delta \G{^\theta_\theta} [h,x) -8 \pi  \delta \T{^\theta_\theta}[h,x) - 8 \pi
\int \mathcal{K}_1 (x-x') \delta \T{^\theta_\theta} [\xi,x') d^4x' = 
\ttau{^\theta_\theta}[g,x)
\ee
which reads,
\bea \label{eq:thetatheta1}
& &[ [e^{-2\nu} \partial_t^2 - \frac{e^{-2 \lambda}}{2} \partial_r^2 -
 \frac{1}{2}
e^{-2 \lambda} ( \frac{2}{r} \nu' - 2\lambda' + \frac{1}{r}- 2 e^{2 \lambda}
\partial_r) \partial_r - 8 \pi \Gamma_1 p ] K \nonumber \\
& & [ - e^{-2 \lambda} ( \partial_r - \frac{1}{r} - \lambda') ( \partial_r +
2 \nu') + \frac{1}{2} e^{-2 \nu} \partial_t^2 + \frac{1}{2} e^{-2 \lambda}
\partial_r^2 + \frac{e^{-2 \lambda}}{2} ( \frac{2}{r} + \nonumber \\
& &  3 \nu' - \lambda') \partial_r - 8 \pi p - 4 \pi \Gamma_1 p ] H_0
+ 8 \pi \mathcal{K}_1(\gamma) [ \Gamma_1 p \frac{e^{- \lambda}}{r^2} \partial_r
+ \frac{p'}{r^2} e^{-\lambda} ] W +  \nonumber \\
& & 8 \pi \mathcal{K}_1(\gamma) \Gamma_1 p \frac{l(l+1)}{r^2} V] P_l
(\cos \theta) =
 \delta_s p_{\tilde{l}}(r) e^{i \omega t} P_{\tilde{l}}(\cos \theta) 
\eea
From (\ref{eq:tr1}) 
\bea \label{eq:W}
W(r,t) P_l (\cos \theta) & = & \frac{r^2 \gamma e^{2 \nu - 
\lambda}}{\tilde{\mathcal{K}}(\gamma) 
( \epsilon + p)} [ \{ - e^{-\lambda + \nu} ( \partial_r + 2 \nu') H_0 P_l(\cos
\theta) 
+ e^{- \lambda + \nu} \partial_r K P_l (\cos \theta) \}   \nonumber \\
& & + 8 \pi e^{\nu - \lambda} (\epsilon+ p)
\delta_s v_{\tilde{l}}(r) e^{i \omega t} P_{\tilde{l}} \cos{\tilde{\theta}} ]
\eea
Substituting this in equation (\ref{eq:tt1}),
\bea \label{eq:V}
V(r,t) P_l (\cos \theta) & = & \frac{r^2}{l(l+1) (\epsilon + p)} \{ -
 [ -e^{-2 \lambda} 
\partial_r^2  - e^{-2 \lambda} ( \frac{3}{r} - \lambda' ) \partial_r + 
[\frac{1}{2 r^2} ( l-1)(l+2) + 8 \pi (\epsilon+ p) ] + \nonumber \\
& & \tilde{\mathcal{K}}_0 (\gamma) 8 \pi ( \frac{e^{- \lambda}}{r^2} 
( (\epsilon+ p) \partial_r + \epsilon' )) \frac{r^2 \gamma e^{2 \nu - 
\lambda}}{ \tilde{\mathcal{K}}(\gamma) 
(\epsilon + p) } e^{- \lambda + \nu } \partial_r ] \} K P_l(\cos \theta)
\nonumber \\
& & + [  \frac{e^{-2 \lambda}}{r} \partial_r - e^{- 2 \lambda}
 ( \frac{1}{r^2} - 
\frac{2}{r} \lambda' + \frac{ e^{2 \lambda}}{2 r^2} l (l+1) ) + 4 \pi (\epsilon
+ p) + \nonumber \\
& & \tilde{\mathcal{K}}_0(\gamma) 8 \pi ( \frac{e^{-\lambda}}{r^2}( ( 
\epsilon+p) \partial_r + \epsilon' )) \frac{ r^2  \gamma e^{2 \nu - 
\lambda}}{ \tilde{\mathcal{K}}(\gamma)
 (\epsilon+ p)} ( \partial_r + 2 \nu') ] H_0  P_l (\cos \theta)   \nonumber \\
& &+ ( 8 \pi )^2 \tilde{\mathcal{K}}_0(\gamma) ( \frac{e^{- \lambda}}{r^2}
 (\epsilon+ p)
\partial_r + \epsilon') r^2 \gamma \frac{e^{ 3 \nu - 2 \lambda}}{
 \tilde{\mathcal{K}}(\gamma)} (\delta_s v(r)  - 
\delta_s \epsilon_{\tilde{l}}) (r) e^{i \omega t} P_{\tilde{l}} 
(\cos \tilde{\theta})
\eea

Using the above in equation (\ref{eq:rr1}), 
\bea \label{eq:fff1}
& & [\{e^{-2 \nu} \partial_t^2 + e^{-2 \lambda} ( \frac{1}{r} - \nu')
 \partial_r - 8 \pi \Gamma_1 p + \frac{1}{r} ( l-1) (l-2) \} - 8 \pi 
\tilde{\mathcal{K}}_1 [\Gamma_1 p e^{- \lambda}{r^2} \partial_r + \nonumber \\
& & \frac{p'}{r^2} e^{- \lambda} ] 
\frac{r^2 \gamma e^{3 \nu - 2\lambda}}{\tilde{\mathcal{K}}(\gamma)
 (\epsilon+ p)}
\partial_r + 8 \pi \tilde{\mathcal{K}}_0 (\gamma) (  \frac{e^{- \lambda}}{r^2}
 ( (\epsilon+ p) \partial_r + \epsilon' ) \frac{r^2 \gamma e^{3 \nu -
 2 \lambda}}{
\tilde{\mathcal{K}}(\gamma) ( \epsilon + p)} \partial_r ] K P_l \cos{\theta} 
 \nonumber \\
& &+ [ \{ \frac{e^{- 2 \lambda}}{r} + \frac{1}{r^2} ( e^{- 2 \lambda} -1) 
- \frac{1}{r}(l-1)(l+2) + 4 \pi \Gamma_1 p - e^{- \lambda + \nu} ( \partial_r
+ 2 \nu') + \frac{e^{-2 \lambda}}{r} \partial_r - e^{-2 \lambda} ( \frac{1}{r^2}
- \frac{2}{r} \lambda' \nonumber \\
& & + \frac{e^{2 \lambda}}{2 r^2} l(l+1) ) + 4 \pi ( 
\epsilon+ p) + \tilde{\mathcal{K}}_0 ( \gamma) 8 \pi ( \frac{e^{- 
\lambda}}{r^2}( (\epsilon + p) \partial_r + \epsilon') \frac{r^2 \gamma 
e^{ 2\nu - \lambda}
(\partial_r + 2 \nu')}{ \tilde{\mathcal{K}} (\gamma) ( \epsilon+ p) }
 (\partial_r + 2 \nu')] H_0 P_l \cos \theta \nonumber \\
& & = - (8 \pi)^2 \{ \tilde{\mathcal{K}}_1 (\gamma) e^{-\lambda}{r^2}
 ( \Gamma_1 p 
\partial_r + p')  + \tilde{\mathcal{K}}_0(\gamma) ( \frac{e^{-\lambda}}{r^2} 
( \epsilon + p) \partial_r + \epsilon') \} \frac{r^2 \gamma e^{3 \nu - 2 
\lambda}}{\tilde {\mathcal{K}}(\gamma)} \delta_s v_{\tilde{l}}(r) 
e^{i \omega t} 
P_{\tilde{l}}( \cos \tilde{\theta}) - \nonumber \\
& & (\delta_s \epsilon_{\tilde{l}}(r) + \delta_s p_{\tilde{l}}(r) )
 e^{-i \omega t} P_{\tilde{l}} (\cos \tilde{\theta}) 
\eea
which can be written in compact form as,  
\be \label{eq:fff2}
\{X_l^1 K(r,t) + Y_l^1 H_0(r,t)\} P_l \cos \theta =  
  \{(-F(r) \delta_s v_{\tilde{l}}(r)  -  \delta_s \epsilon_{\tilde{l}}(r) +
 \delta_s p_{\tilde{l}}(r) ) e^{-i \omega t} \} P_{\tilde{l}} 
(\cos \tilde{\theta}) 
\ee
The coefficients $X_l^1, Y_l^1, F(r) $ can be matched  with the terms in the
above equations.

We also have from equation (\ref{eq:the11}),
\bea \label{eq:fff3}
 & & \{e^{- 2 \nu} \partial_t^2 - e^{-2 \lambda}{2} \partial_r^2 - \frac{1}{2}
 e^{- 2 \lambda} ( \frac{2}{r} \nu' - 2 \lambda' + \frac{1}{r} -
 2 e^{ 2 \lambda} \partial_r ) \partial_r -  8 \pi \Gamma_1 p + 
8 \pi \tilde{\mathcal{K}_1}(\gamma)
[\Gamma_1 p \frac{e^{- \lambda}}{r^2} \partial_r + \frac{p'}{r^2} 
e^{- 2 \lambda} ] \frac{r^2 \gamma e^{ 3 \nu - 2\lambda}}{\tilde{K}(\gamma)
 ( \epsilon + p) } e^{- \lambda + \nu} \partial_r \nonumber \\
& & - 8 \pi \frac{\mathcal{K}_1(\gamma)}{(\epsilon+ p) } [  - e^{- 2 \lambda}
 \partial_r^2  - e^{ -2 \lambda} 
(\frac{3}{r} - \lambda' ) \partial_r + [ \frac{1}{2 r^2} (l-1)(l+2)
+ 8 \pi ( \epsilon+ p) ] + \tilde{\mathcal{K}}_0(\gamma) 8 \pi (
 \frac{e^{- \lambda}}{r^2} 
( ( \epsilon+ p)) \partial_r + \epsilon') \frac{r^2 \gamma 
e^{3 \nu - 2 \lambda} }{\tilde{K}(\gamma) (\epsilon+ p) } \partial_r ] \}K
\nonumber \\
& & 
+  \{ [ - e^{-2 \lambda} ( \partial_r - \frac{1}{r} - \lambda') (\partial_r
+ 2 \nu') + \frac{1}{2} e^{- 2 \nu} \partial_t^2 + \frac{1}{2} e^{-2 \lambda}
\partial_r^2 + \frac{e^{-2 \lambda}}{2} ( \partial{2}{r} + 3 \nu' - \lambda')
\partial_r  - 8 \pi p - 4 \pi \Gamma_1 p ] - \nonumber \\
& & 8 \pi \mathcal{K}_1(\gamma) [\Gamma_1 p
\frac{e^{-\lambda}}{r^2} \partial_r + \frac{p'}{r^2} e^{-2 \lambda} ] 
{r^2} \gamma \frac{e^{3 \nu - 2\lambda}}{\tilde{K}(\gamma) (\epsilon+ p)}
\{ e^{- \lambda + \nu}(\partial_r + 2 \nu' ) \} 
+ [ e^{-2 \lambda}{r} \partial_r - e^{- 2 \lambda} ( \frac{1}{r^2} -
 \nonumber \\
 & &  \frac{2 \lambda'}{r} + 
\frac{e^{2 \lambda}}{2 r^2} l (l+1) ) + 4 \pi ( \epsilon+p) +
\tilde{\mathcal{K}}_0 (\gamma) 8 \pi ( \frac{e^{-\lambda}}{r^2} ((\epsilon+ p)
 \partial_r 
+ \epsilon' ) \frac{r^2 \gamma e^{2 \nu - \lambda}}{\tilde{K}(\gamma) 
(\epsilon + p)} ( \partial_r + 2 \nu') ] \} H_0 = \nonumber \\
& &  \{( \delta_s p_{\tilde{l}}(r)) + \delta_s 
\epsilon_{\tilde{l}}(r) \} e^{ i \omega t} 
P_{ \tilde{l}} (\cos \tilde{\theta}) - ( 8 \pi)^2 
\{ \tilde{\mathcal{K}}_1(\gamma)
 ( \Gamma_1 p \frac{e^{-\lambda}}{r^2} \partial_r + 
\frac{p'}{r^2} e^{-2 \lambda} + \tilde{\mathcal{K}}_0(\gamma) 
( \frac{e^{-\lambda}}{r^2}
(\epsilon+p) \partial_r + \epsilon') \} \nonumber \\
& & \frac{r^2 \gamma e^{3 \nu - 2 \lambda}}
{ \tilde{\mathcal{K}}(\gamma) } \delta_s v_{\tilde{l}}(r) e^{i \omega t}
 P_{\tilde{l}} ( \cos \tilde{\theta}) 
\eea
which is  written in compact form as,
\be \label{eq:fff4}
 \{ X_l^2 K(r,t) + Y_l^2 H_0(r,t) \} P_l \cos \theta =
   (\delta_s p_{\tilde{l}}(r)  +
 \delta_s \epsilon_{\tilde{l}}(r)  
 - F(r) \delta_s v_{\tilde{l}}) e^{i \omega t} P_{\tilde{l}}
 \cos (\tilde{\theta}) 
\ee 
 The coefficients $X_l^2, Y_l^2 , F(r)$ can be matched with the terms in the
above equation.
From equations (\ref{eq:fff1}), (\ref{eq:fff2}) and (\ref{eq:fff3}), 
(\ref{eq:fff4}) one can evaluate $H_0$ and $K$ while $W$ and $V$ are given by
(\ref{eq:W}) and (\ref{eq:V}) respectively. This will need
numerical methods to solve them for $H_0$ and $K$, we have presented the
 analytical
form of the basic equations here. However from the expressions it is clear
that these will be expressed in terms of some combination of
$\delta_s\epsilon_{\tilde{l}}(r,t), \delta_s p_{\tilde{l}}(r,t), 
\delta_s v_{\tilde{l}}(r,t) $. Thus it is the same term which induces the polar
perturbations with characteristic modes, that can be analysed in future work
to understand the new scales in compact matter which are yet untouched. We
expect new structural properties as well as new phenomena at these scales to
emerge as a result of our speculations. 
\section{Conclusion}
In this article we have solved the inhomogeneous sourced Einstein's equations
 for
a spherically symmetric non-rotating relativistic star with cold
dense matter fluid in a near equilibrium configuration. 
We have given a first principles account of  perturbations
induced at mesoscopic scales in a relativistic star, due to internal sources
 in the cold dense 
compact matter. The radial perturbations are of significance to study stability
properties, while the polar perturbations connect
to interests in gravitational waves. The significance of the induced 
perturbations lies in realizing that they are different from the implicit
perturbations that arise from homogenous solutions of the Einstein's equation.
Here the inhomogeneous source term is of central importance to characterize the
internal structure and mechanism at mesoscopic scales for the dense compact 
matter. Hence with our approach and formulations, we are trying to build up
a new theory for sub-hydro mesoscopic scale physics for the exotic fluids that
has not yet been touched upon. The mesoscopic scales  show their typical 
behaviour characterized by the modes of oscillations both for deterministic
and stochastic cases. Through the basic new mathematical developments  that
we do in this article,  new features of the astrophysical system
can be touched upon for further investigations. In future work we will move
 towards
characterizing turbulence in the superfluid state of the exotic matter, 
with emphasis on the stochastically induced perturbations for qualitative as
well as quantitative analysis. This will need further formulations to relate
the realistic configurations that provide us with a new theoretical base for
 possible observations. It is expected that this  will give access 
 to dynamical effects and interior structure of the rotating and 
gravitationally radiating  compact star
 configuration with more refined features than are currently under focus. 
 The work done in this article
is a refinement of the basic new principles in order to establish a 
sub-hydro meso-scale theory for exotic dense matter present .
We expect that the extreme physical conditions existing inside compact
stars  show up distinct results in the 
sub-hydro meso scales which connect the microscales to macroscopic
scales while retaining their own specific dynamical and structural properties.
This also gives way for the intermediate scale physics in extended non-local
 strong gravity regions. 

\textbf{Data Sharing Statement}: No data was generated in the work related to
 this article.

\section*{Acknowledgements}

The work done in this article is funded through grant no. DST/WoSA/PM-3/2021.
Dept of Science and Technology, India. 

\section*{Appendix}
\begin{center}
\textbf{Review of polar perturbations in spherically symmetric star}
\end{center}

The polar perturbations are even-parity perturbations on the 2-sphere of the
metric. The perturbed metric in Regge Wheeler gauge then is given by
\[
h_{\alpha \beta} =
\begin{bmatrix} e^{2 \nu} H_0 Y_{lm} && H_1 Y_{lm} && 0 && 0 \\
Sym & & e^{2 \lambda } H_2 Y_{lm} &&  0 && 0 \\
Sym && Sym && r^2 K Y_{lm} && 0 \\
Sym & & Sym && Sym && r^2 K \sin^2 \theta Y_{lm}
\end{bmatrix}
\]
Eventually the independent metric variables are commonly taken to be
$H_0, H_1,$ and $K$ as it turns out that $H_2 =H_0$ using the Einstein's
tensors $ \delta G^\theta_\theta - \delta G^\phi_\phi =0 $.
\bea
0 &= & \delta G^\theta_\theta - \delta G^\phi_\phi =(H_2 - H_0)
\frac{1}{2 r^2}(\partial_\theta^2 - \cot \theta \partial_\theta -
\frac{1}{\sin^2 \theta}\partial^2_\phi ) Y_{lm}(\theta, \phi)
\eea
or $ H_0 = H_2 $.
Also from
\be
\delta G_{r \theta} = 0
\ee
\bea
& & e^{-2 \nu} \partial_t H_1 = H_0' + 2 \nu' H_0 - K'
\eea
The perturbed fluid four-velocity with $Y_{lm}$ replaced by the condition
$m=0$.
\bea
\delta u^t &=& - e^{\nu} [ 1- \frac{1}{2} H_0 P_l(\cos \theta)] \\
\delta u^r & = & -\frac{e^{-(\nu+\lambda}}{r^2} W_{,t} P_l (\cos \theta) \\
\delta u^\theta  & = & e^{\nu}\frac{1}{r^2} V_{,t} \partial_\theta
P_l(\cos \theta)\\
\delta u^\phi & = & 0
\eea
The Lagrangian change in  number density of baryons  is $\Delta n $, then
\be
\Delta n/n = \{-\frac{e^{-\lambda}}{r^2} W' - \frac{l(l+1)}{r^2}V + \frac{1}{2}
H_2 + K \} P_l (\cos \theta).
\ee
The corresponding Eulerian changes in energy density and pressure are
\bea
\delta \epsilon & = & (\epsilon+p) (\Delta n/n) - \epsilon' e^{-\lambda} W
P_l (\cos \theta) \\
\delta p & = & \Gamma p (\Delta n/n) - p' \frac{e^{-\lambda}}{r^2} W
P_l(\cos \theta) .
\eea
The Eulerian changes in the stress-energy tensor are
\bea
& & \delta T^t_t = - \delta \epsilon, \delta T^r_r = \delta T^\theta_\theta
= \delta T^\phi_\phi = \delta p , \\
& &  \delta T^r_t  = (\epsilon+p) u_t \delta u^r, \delta T^t_r = (\epsilon+ p)
 u_r \delta u^t, \\
& & \delta T^\theta_t = (\epsilon +p) u_t \delta u^\theta ,
\delta T^t_\theta = (\epsilon+ p) u_\theta \delta u^t
\eea
Rest of the components of $\delta T^\alpha_\beta $ vanish.
\end{document}